\documentclass[conference]{IEEEtran}
\IEEEoverridecommandlockouts

\usepackage[T1]{fontenc}

\usepackage{booktabs} %
\usepackage{hyperref}
\usepackage{cite}
\usepackage[10pt]{moresize}
\usepackage{graphicx}
\usepackage{dsfont}
\usepackage{latexsym}
\usepackage{csquotes}
\usepackage{tcolorbox}
\usepackage{listings}
\usepackage{float}
\usepackage{multirow}
\usepackage[scaled]{helvet}
\usepackage[noend]{algpseudocode}
\usepackage{mathrsfs}
\usepackage[linesnumbered, ruled, lined]{algorithm2e}  %
\usepackage{mathpartir}
\usepackage{mathtools}
\usepackage{stmaryrd}
\usepackage{textcomp} 
\usepackage{bbm}
\usepackage{verbdef}
\usepackage{xspace}
\usepackage{verbatim}
\usepackage{lipsum}
\usepackage{wrapfig}

\usepackage[shortlabels]{enumitem}
\usepackage{pifont}
\usepackage{varwidth}
\usepackage{xpatch}
\usepackage{multirow}
\usepackage[dvipsnames]{xcolor}
\usepackage[export]{adjustbox}
\usepackage{caption}
\usepackage{subcaption}
\usepackage{setspace}
\usepackage[capitalize]{cleveref}
\usepackage{makecell}
\usepackage{dashrule}
\usepackage{comment}
\usepackage{arydshln}
\usepackage{ifsym}
\usepackage{mdframed}
\usepackage{todonotes}

\usepackage[varqu]{zi4}

\usepackage{flushend}

\usepackage{tikz}
\usetikzlibrary{calc,positioning,math, arrows, shapes}

\AtBeginDocument{%
  \providecommand\BibTeX{{%
    \normalfont B\kern-0.5em{\scshape i\kern-0.25em b}\kern-0.8em\TeX}}}

\setlist[itemize]{topsep=0pt,itemsep=-1ex,partopsep=1ex,parsep=1ex,itemindent=0pt,leftmargin=8pt}
\definecolor[named]{ACMBlue}{cmyk}{1,0.1,0,0.1}
\definecolor[named]{ACMYellow}{cmyk}{0,0.16,1,0}
\definecolor[named]{ACMOrange}{cmyk}{0,0.42,1,0.01}
\definecolor[named]{ACMRed}{cmyk}{0,0.90,0.86,0}
\definecolor[named]{ACMLightBlue}{cmyk}{0.49,0.01,0,0}
\definecolor[named]{ACMGreen}{cmyk}{0.20,0,1,0.19}
\definecolor[named]{ACMPurple}{cmyk}{0.55,1,0,0.15}
\definecolor[named]{ACMDarkBlue}{cmyk}{1,0.58,0,0.21}

\makeatletter
\def\arcr{\@arraycr}
\makeatother

\makeatletter
\newcommand{\mtmathitem}{%
\xpatchcmd{\item}{\@inmatherr\item}{\relax\ifmmode$\fi}{}{\errmessage{Patching of \noexpand\item failed}}
\xapptocmd{\@item}{$}{}{\errmessage{appending to \noexpand\@item failed}}}
\makeatother

\definecolor{shadecolor}{gray}{1.00}
\definecolor{ddarkgray}{gray}{0.75}
\definecolor{darkgray}{gray}{0.30}
\definecolor{light-gray}{gray}{0.87}

\newcommand{\etal}{\emph{et~al.}\xspace}

\definecolor{pblue}{rgb}{0.13,0.13,1}
\definecolor{pgreen}{rgb}{0,0.5,0}
\definecolor{pred}{rgb}{0.9,0,0}
\definecolor{pgrey}{rgb}{0.46,0.45,0.48}

\definecolor{ckeyword}{HTML}{7F0055}
\definecolor{ccomment}{HTML}{3F7F5F}
\definecolor{cnumber}{HTML}{2A0099}
\definecolor{darkgreen}{HTML}{008000}

\definecolor{clr-background}{RGB}{255,255,255}
\definecolor{clr-text}{RGB}{0,0,0}
\definecolor{clr-string}{RGB}{163,21,21}
\definecolor{clr-namespace}{RGB}{0,0,0}
\definecolor{clr-preprocessor}{RGB}{128,128,128}
\definecolor{clr-keyword}{RGB}{0,0,255}
\definecolor{clr-type}{RGB}{43,145,175}
\definecolor{clr-variable}{RGB}{0,0,0}
\definecolor{clr-constant}{RGB}{111,0,138} %
\definecolor{clr-comment}{RGB}{0,128,0}

\lstdefinestyle{VS2017}{
	backgroundcolor=\color{clr-background},
	basicstyle=\color{clr-text}, %
	stringstyle=\color{clr-string},
	identifierstyle=\color{clr-variable}, %
	commentstyle=\color{clr-comment},
	directivestyle=\color{clr-preprocessor}, %
	keywordstyle=\color{clr-type},
	keywordstyle={[2]\color{clr-constant}}, %
  tabsize=2,
}
\lstdefinelanguage{JavaScript}{
  keywords={typeof, new, true, false, catch, function, return, null, catch, switch, var, if, in, while, do, else, case, break, const, let, async, await},
  keywordstyle=\color{blue}\bfseries,
  ndkeywords={class, export, boolean, throw, implements, import, this},
  ndkeywordstyle=\color{darkgray}\bfseries,
  identifierstyle=\color{black},
  sensitive=false,
  comment=[l]{//},
  morecomment=[s]{/*}{*/},
  commentstyle=\color{purple}\ttfamily,
  stringstyle=\color{red}\ttfamily,
  morestring=[b]',
  morestring=[b]"
}

\lstdefinelanguage{Kotlin}{
  comment=[l]{//},
  commentstyle={\color{gray}\ttfamily},
  emph={filter, first, firstOrNull, forEach, lazy, map, mapNotNull, println},
  emphstyle={\color{OrangeRed}},
  identifierstyle=\color{black},
  keywords={!in, !is, abstract, actual, annotation, as, as?, break, by, catch, class, companion, const, constructor, continue, crossinline, data, delegate, do, dynamic, else, enum, expect, external, false, field, file, final, finally, for, fun, get, if, import, in, infix, init, inline, inner, interface, internal, is, lateinit, noinline, null, object, open, operator, out, override, package, param, private, property, protected, public, receiveris, reified, return, return@, sealed, set, setparam, super, suspend, tailrec, this, throw, true, try, typealias, typeof, val, var, vararg, when, where, while},
  keywordstyle={\color{NavyBlue}\bfseries},
  morecomment=[s]{/*}{*/},
  morestring=[b]",
  morestring=[s]{"""*}{*"""},
  ndkeywords={@Deprecated, @JvmField, @JvmName, @JvmOverloads, @JvmStatic, @JvmSynthetic, Array, Byte, Double, Float, Int, Integer, Iterable, Long, Runnable, Short, String, Any, Unit, Nothing},
  ndkeywordstyle={\color{BurntOrange}\bfseries},
  sensitive=true,
  stringstyle={\color{ForestGreen}\ttfamily},
}
\definecolor{bluekeywords}{rgb}{0.13,0.13,1}
\definecolor{greencomments}{rgb}{0,0.5,0}
\definecolor{turqusnumbers}{rgb}{0.17,0.57,0.69}
\definecolor{redstrings}{rgb}{0.5,0,0}
\lstdefinelanguage{WebAssembly}{
  sensitive=true,
  otherkeywords={},
  morekeywords=[1]{i32,f32,i64,f64},
  keywordstyle={[1]\color{violet}},
  morekeywords=[2]{0},
  keywordstyle={[2]\color{violet}},
  morekeywords=[3]{add,const}
  keywordstyle={[3]\color{bluemunsell}},
  morekeywords=[4]{},
  keywordstyle={[4]\color{candypink}},
  morekeywords=[5]{module, func, param, result, global, get_global, mut, set_global, export, import, memory, data, get_local, set_local, elem, table, call,call_indirect, type},
  keywordstyle={[5]\color{bluekeywords}},
  morekeywords=[6]{=,;},
  keywordstyle={[6]\color{britishracinggreen}},
  morekeywords=[7]{(,),[,],.},
  keywordstyle={[7]\color{black}},
  numberstyle=\tiny\color{black},
  rulecolor=\color{black},
  morecomment=**[l][\itshape\color{greencomments}]{;;},
}

\usepackage{lipsum}

\usepackage{soul}
\usepackage[justification=centering]{caption}
\definecolor{lightgray}{rgb}{.9,.9,.9}
\definecolor{darkgray}{rgb}{.4,.4,.4}
\definecolor{purple}{rgb}{0.65, 0.12, 0.82}

\lstdefinelanguage{JavaScript}{
	keywords={break, case, catch, continue, debugger, default, delete, do, else, false, finally, for, function, if, in, instanceof, new, null, return, switch, this, throw, true, try, typeof, var, void, while, with},
	morecomment=[l]{//},
	morecomment=[s]{/*}{*/},
	morestring=[b]',
	morestring=[b]",
	ndkeywords={class, export, boolean, throw, implements, import, this},
	keywordstyle=\color{blue}\bfseries,
	ndkeywordstyle=\color{darkgray}\bfseries,
	identifierstyle=\color{black},
	commentstyle=\color{purple}\ttfamily,
	stringstyle=\color{red}\ttfamily,
	sensitive=true
}

\usepackage{xurl}        %
\usepackage{hyperref}    %
\hypersetup{breaklinks=true} %

\begin{document}

\title{Towards Analyzing N-language Polyglot Programs}

\author{
	\IEEEauthorblockN{
		Jyoti Prakash,
		Abhishek Tiwari\IEEEauthorrefmark{1},
		Mikkel Baun Kjærgaard
	}
	\IEEEauthorblockA{
		SDU Software Engineering, The Mærsk McKinney Møller Institute\\
		University of Southern Denmark\\
		Odense, Denmark\\
		\{jyp, abti, mbkj\}@mmmi.sdu.dk
	}
\thanks{\IEEEauthorrefmark{1}Corresponding author.}
}

\maketitle

\begin{abstract}
    
Polyglot programming is gaining popularity as developers integrate multiple programming languages to harness their individual strengths. With the recent popularity of platforms like GraalVM and other multi-language runtimes, creating and managing these systems has become much more feasible. However, current research on analyzing multilingual programs mainly focuses on two languages, leaving out the increasing complexity of systems that use three or more. For example, modern web systems often link JavaScript, WebAssembly, and Rust within the same execution chain. This paper envisions the landscape of software systems with three-language polyglot communication. We identify fundamental challenges in analyzing them and propose a conceptual roadmap to advance static analysis techniques to address them. Our vision aims to stimulate discussion and inspire new research directions toward scalable, language-agnostic analysis frameworks for next-generation polyglot systems.

\end{abstract}


\section{Introduction}

 Recent years have seen a growing trend toward polyglot programming. Developers integrate multiple programming languages to harness specific capabilities, e.g., high performance from C/C++ or robust type systems in Java. Frameworks such as GraalVM, Android NDK, and  WebView enable these combinations, leading to polyglot programs that combine diverse runtime environments~\cite{davidlo2016polyglot, mlpolyglotstudy, stackexchange, polyglotstudy2, rcpp, rpy, adlib2019ryu, hybridDroid, JuCify2022ICSE, hydridDroidICSE2019}. 

While these environments enable developers to combine the strengths of multiple languages, they also introduce new sources of complexity. Polyglot programming presents several challenges~\cite{davidlo2016polyglot, mlpolyglotstudy, polyglotstudy2, LiLiVulnerabilityMultilingual}. For example, integrating code written in different languages can be complex and error-prone, particularly when dealing with incompatible data types or calling conventions. Additionally, debugging and profiling polyglot programs can be challenging, as traditional tools and techniques may not work across language boundaries. Furthermore, ensuring the correctness and security of polyglot programs can be difficult, as vulnerabilities in one language may be exploited by code written in another language. 

Fortunately, these challenges can be addressed through static analysis. Existing research has made progress on analyzing polyglot programs with two-language interoperability~\cite{hydridDroidICSE2019, JuCify2022ICSE, iwandroid, LuDroid-Journal, tiwari2019ludroid}: for instance, Java–C (JNI), Java–JavaScript (WebView), or Java-Python (Jython). Yet, real-world systems have started to exceed this boundary. Android hybrid applications, for example, combine Java, JavaScript, and WebAssembly. Similarly, emerging web systems integrate JavaScript, WebAssembly, and Rust. These settings embody three-language polyglot programs, in which communication chains span multiple language boundaries.

While two-language analyses have advanced, they fail to scale naturally due to their reliance on isolated semantic boundaries. Extending them to three or more languages, these assumptions collapse: dataflows may traverse multiple runtime abstractions, creating mutual dependencies and cyclic interactions that traditional frameworks cannot capture. We argue that this is not a mere engineering inconvenience; it is a theoretical and methodological gap. A three-language system is not \emph{just one more edge} in a communication graph; it introduces new semantic phenomena—multi-hop dependencies, cyclic shared variables, and emergent dataflow transformations—that no existing two-language model can capture.

This paper explores this uncharted territory, arguing that the future of software analysis must embrace a new paradigm: \emph{n-language polyglot system}.
We present a conceptual framework that characterizes how three-language systems communicate, identify open challenges in static analysis for such systems, and propose a research roadmap toward scalable, summary-based analysis that spans multiple language boundaries.

To summarize, this work makes the following contributions:
\begin{itemize}
	\item We characterize the landscape of three-language polyglot communication models.
	\item We identify key challenges for static analysis, including inter-language dataflow propagation and mutual fixed-point computation.
	\item We propose a visionary roadmap toward unified, language-agnostic frameworks for analyzing polyglot software systems beyond two languages.
\end{itemize}
Our vision aims to spark a discussion within the research community about how future analyses should evolve in an increasingly polyglot world, where language boundaries blur, and the real challenge lies not in analyzing one language at a time, but in understanding the interactions among them.

\section{Related Work}
To the best of our knowledge, there exists no work on three-language polyglot communication. However, two-language polyglot communication has been studied in the past. Lee~\etal~\cite{ryu-semantic-summary} proposed a modular semantic summary extraction technique for analyzing Java-C/C++ interoperations. Mon\"et~\etal~\cite{Monat-Python-SAS} used abstract interpretation for analyzing Python and C programs. JN-SAF~\cite{weiJNSAF-CCS2018} by Wei~\etal used a function summary-based bottom-up dataflow analysis for analyzing Java Native applications. Buro~\etal proposed an abstract interpretation-based framework~\cite{buro2020abstractmultlingual} for analyzing multilingual programs by combining analysis of while and expression language based on abstract interpretation to combine a while and expression language analyses. Youn~\etal (2023) proposed a declarative multilingual analysis~\cite{declarativeMutlilingual} embedded in the CodeQL framework to fuse the data flows at the bridge interfaces. Negrini~\etal~(2023)~\cite{Negrini2023} proposed a generic framework, \emph{Lisa}, for multilingual static analysis based on a common intermediate representation providing the necessary infrastructure for the developers to customize the analysis for each language and combine those for a multilingual analysis. Prakash~\etal~\cite{PRAKASH2025103278} proposed a unifying framework for pointer analysis in two-language polyglot programs. Liu~\etal~\cite{Reunify} proposed a multilingual analysis framework for analyzing React-Native applications. Tiwari~\etal~\cite{iwandroid} proposed a multilingual analysis framework for analyzing Android WebView programs.

\section{Landscape of Multilingual Programs}
Multilingual programs consist of more than one language, with one or more languages acting as hosts and others as guests. The host initiates program execution, typically through a main method or event-driven entry points. It triggers the execution of the guest language, either through events or direct function calls. After the host calls a guest function, interactions can occur back and forth between the two languages. In the following, we describe three communication models of multilingual programs.

\subsection{Multilingual Programs Running on Different Runtime Environments (Model I)}
\begin{figure}[tb]
    \centering
    \includegraphics[width=\columnwidth]{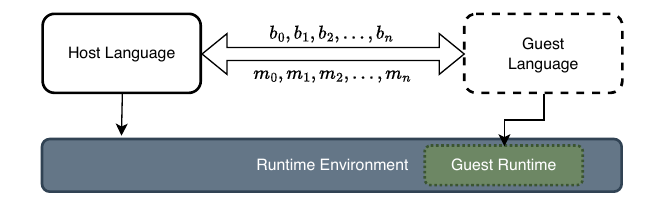}
    \caption{Multilingual programming through different runtime environments}
    \label{fig:multilingual-programs-bridge}
\end{figure}

In this case, one language typically assumes the role of the host language, as shown in~\cref{fig:multilingual-programs-bridge}. The host initiates and controls the program’s execution, while the other language serves as the guest, which the host invokes to perform specific tasks. The interaction between these two languages is facilitated by a foreign-function interface (FFI), which acts as a bridge, enabling seamless communication and data exchange between the host and guest languages. The guest runtime environment is embedded within the host and invoked either through the foreign-function interface or when the guest calls a function shared between the host and guest, ensuring efficient collaboration between the two environments.

\subsection{Multilingual Programs through Common IR (Model II)}

\begin{figure}[tb]
	\centering
	\includegraphics[width=\columnwidth]{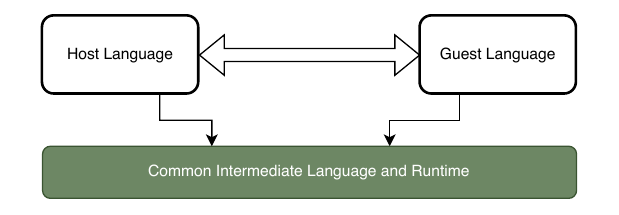}
	\caption{Multilingual through common IR}
	\label{fig:multilingual-programs-ir}
\end{figure}

In this scenario, a common intermediate representation (IR) is employed, as shown in~\ref{fig:multilingual-programs-ir}, where a single runtime environment executes both the host and guest languages. This approach enables seamless integration between the languages by compiling both the host and guest languages into a shared intermediate language that the runtime understands and processes. By using a common IR, the runtime eliminates many of the complexities associated with language boundaries, such as differing memory models or calling conventions. This seamless integration ensures consistent performance and interoperability, allowing both languages to leverage shared runtime features, such as garbage collection, type checking, and optimizations while maintaining the distinct semantics of each language.

\subsection{Multilingual Programs through Shared Libraries (Model III)}

\begin{figure}[tb]
	\centering
	\includegraphics[width=\columnwidth]{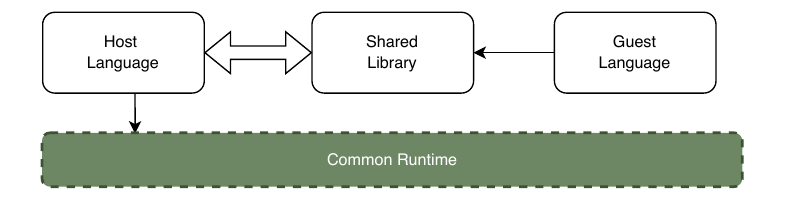}
	\caption{Multilingual through shared libraries}
	\label{fig:multilingual-programs-shared-libraries}
\end{figure}

An alternative approach to multilingual programming involves compiling the guest language into a library that can be shared with the host language code, shown in~\cref{fig:multilingual-programs-shared-libraries}. In this scenario, the runtime environment for the guest language library may either be embedded within the host language's runtime or rely on a common runtime environment shared between both languages. For instance, Java Native Interface (JNI) allows Java programs to invoke native libraries written in C or C++ by embedding their runtimes directly into the Java execution environment. Similarly, frameworks like the .NET Common Language Runtime (CLR) provide a unified execution environment where multiple languages, such as C\# and F\#, can seamlessly interact by compiling to the same intermediate language. This approach ensures modularity and flexibility, allowing developers to integrate functionalities written in different languages while leveraging either the host language's runtime or a shared runtime infrastructure for efficient execution and interoperability.

\section{Polyglot Models with Three Languages}

In a three-language communication setup, there are two interesting scenarios: (1) there is one host and two guests, and (2) there are two hosts and two guests where one language acts as both host and guest. Consider three languages: A, B, and C. In the first scenario, A could be the host, and B and C would be the guests. An example of this setting would be in Android hybrid apps where Java acts as a host for Javascript (via Model I) and Native Code (via Model III). The second scenario has a potential communication chain, e.g., A is a host for B, and B is a host for C. A potential example of this setting would be in Android hybrid apps where Java acts as the host for Javascript (via Model I), and Javascript acts as the host for WebAssembly (via Model II). 

\begin{figure*}[tb]
\centering
    \begin{subfigure}{\linewidth}
        \centering
    \fbox{$\begin{array}{lll}
        f \in \textsc{Functions}            & := & m_1 \mid m_2 \mid m_3 \mid \cdots \mid m_n \\
        i \in \mathit{Bridge Variable}  & :=  & i_1 \mid \cdots  \mid i_n \\
        v \in \textsc{Variables}            & := & v_1 \mid v_2 \mid v_3 \mid \cdots \mid v_n \\
        g \in \textsc{Fields}               & := & g_1 \mid g_2 \mid g_3 \mid \cdots \mid g_n \\
        s \in \textsc{Statements}           & := & v=\mathbf{new}~T()  \mid i=\mathbf{new}~T() \mid \mathbf{return}~v  \\
                                            &    &  v' = v.m() \mid v' = v.g \mid v'.g = v  \mid \mathbf{if}~(e)~\{ s \} [\mathbf{else}~\{s\}]\\
                                            &    & \mid \mathbf{op}~s_1~s_2 \mid s_1;s_2 \mid \mathbf{while}~(e)~\{s\} \\
                                            &    & \mathbf{eval}(program) \mid \mathbf{asmcall}(program) \\
    \end{array}$}
    \caption{Abstract Language for Programs with Foreign Function Interfaces (FFI)}
    \label{fig:language-model-with-interface-variables-host}
    \end{subfigure}

    \begin{subfigure}{0.75\linewidth}
        \centering
        \begin{adjustbox}{width=\columnwidth}
        \fbox{$\begin{array}{lll}
            mod \in \textsc{Modules} & := & mod_1 \mid mod_2 \mid mod_3 \mid \cdots \mid mod_n \\
            p \in \textsc{Procedures} & := & p_1 \mid p_2 \mid p_3 \mid \cdots \mid p_n \mid exportedProcedures \\
            ep \in \textsc{Exported Procedures} & := & \textbf{export}~ep_1 \mid \textbf{export}~ep_2 \mid \textbf{export}~ep_3 \mid \cdots \mid \textbf{export}~ep_n \\
            l \in \textsc{Local Variables} & := & l_1 \mid l_2 \mid l_3 \mid \cdots \mid l_n \\
            g \in \textsc{Global Variables} & := & g_1 \mid g_2 \mid g_3 \mid \cdots \mid g_n \\
            labelstmt \in \textsc{Label Statements} & := & [label]~s \\
            s \in \textsc{Statements} & := & l_1 \leftarrow \textbf{load}~l_2 \mid \textbf{store}~l_1~l_2 \mid \textbf{call}~p \mid \textbf{ret}~l \mid \textbf{br}~label \\
                                      &    & \mathbf{compare}~l_1~l_2 \mid \textbf{branchcond}~label \mid l_1 \leftarrow \textbf{op}~l_2~l_3\\
        \end{array}$}
       \end{adjustbox}
        \caption{\textbf{Abstract Assembly Language}}
        \label{fig:abstract-assembly-language}
    \end{subfigure}

    \caption{MiniDSL for three-language communication}
    \label{fig:language-model}
\end{figure*}

In our work, we focus on the second scenario, i.e., polyglot communication with two hosts and two guests. Instead of using the terms ``host language'' and ``guest language’, we categorize these languages as the entry language, which initiates program execution; the middle language, which acts as both the host and guest; and the bottom language, often embedded through shared libraries. These languages could be integrated in various ways to form a three-language chain. One such approach is for the entry language to invoke the middle language (e.g., via Model I), and the middle language integrates the bottom language via shared libraries (via Model III). These languages may execute on their respective runtimes designed to support them individually.

\Cref{fig:language-model} depicts two abstract languages inspired by the scenarios that can happen in a multilingual environment. For clarity and simplicity and to keep the focus on polyglot communication, many high-level language features are omitted from the syntax, and these languages are presented as imperative/Object-oriented languages. \Cref{fig:language-model-with-interface-variables-host} specifies the abstract language model with specifically marked function calls that facilitate polyglot communication. This language includes constructs for both the host and guest. The constructs, \emph{eval(program)} and \emph{asmcall(program)}, facilitate calling the \emph{program} written in the guest language. The host language can call the guest language program using the \emph{eval} construct. The guest functions called through \emph{eval} are executed in the guest’s runtime. The host language can also call the guest program written intermediate/low-level representation using the \emph{asmcall} construct. The guest functions called via \emph{asmcall} share the same IR as the host and are thus executed in the same runtime as the host. $BridgeVariables$ (denoted by $i$) are special host objects that are designed to expose a set of host functions to the guest. The guest can invoke corresponding host methods using the $BridgeVariables$. 

\begin{figure}[tb]
	\centering
	\includegraphics[width=\columnwidth]{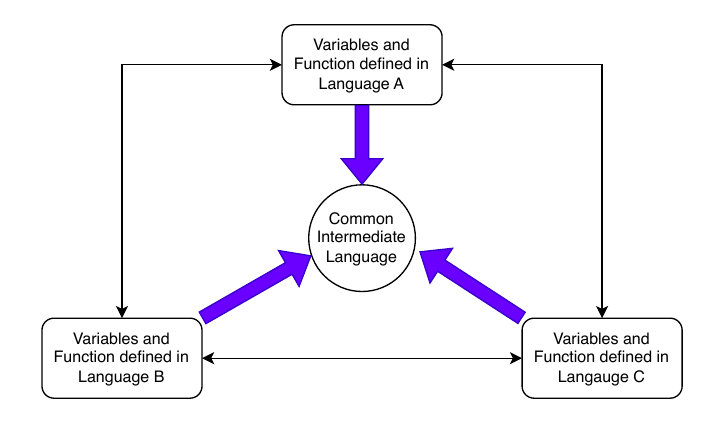}
	\caption{Multilingual through common IR}
	\label{fig:multilingual-programs-ir1}
\end{figure}

\Cref{fig:multilingual-programs-ir1} depicts a polyglot model where three languages share a common intermediate representation.  The languages are compiled into a common intermediate representation (IR) and executed by a common runtime in this model. By using a common IR, the runtime eliminates many of the complexities associated with language boundaries, such as differing memory models or calling conventions.  This seamless integration ensures consistent performance and interoperability, allowing both languages to leverage shared runtime features, such as garbage collection, type checking, and optimizations, while maintaining the distinct semantics of each language. An example of this communication is a polyglot model with Java, Scala, and Kotlin. These languages are nearly interoperable (with appropriate type conversions) because they share the Java bytecode as a common intermediate representation. Consequently, libraries written in any of these languages could be combined, enabling their use as shared libraries across the ecosystem.

\section{Summary based static analysis}
In the following, we present the challenges and pathways for summary-based static analysis, categorized for multilingual programs with a bridge variable and through a common IR.
\subsection{Summary based static analysis with bridge variables}

With the presence of bridged variables in the polyglot programs, the static analysis becomes more complex. As bridged variables are shared between the two languages, the dataflow analysis has to handle the data flow across the language boundaries. One of the examples is to compute the summaries of the functions in the two languages and then merge them to compute the summary of the bridged variables. This is a challenging task as the summaries of the functions in the two languages are incomplete. For example, if a host shares the bridge variable $i$ with a guest, and the guest accesses a host method $i.m()$, then it has first to determine the type of the variable $i$ and then compute the target of the call, which is obtained from the dataflow information available within the host language summary. To give a brief intuition, resolving polyglot communication requires propagating the summaries across the bridge variables and then computing the dataflow information across the language boundaries for each pair of host-guest languages. This has been addressed in a scenario of two-language communion in the prior work~\cite{prakash2023unifyingpointeranalysespolyglot}. Extending this idea to three-language polyglot communication will require additional care to handle the complexities of the three languages.

\paragraph{Challenge: Mutual Fixed Point in incremental propagation of resolved dataflows} Static analysis computes a fixed point over transfer functions defined by program statements. In multilingual programs, this generalizes to mutual fixed points, where separate analyses for each language exchange information at inter-language boundaries until convergence. For two-language systems, this involves iteratively reconciling dataflows between the host and guest languages; for three or more languages, the complexity grows combinatorially, as fixed points must be computed for each pairwise interaction. This significantly increases analysis runtime and poses challenges for scalability.

A key open problem is how to incrementally update only the affected dataflows at language boundaries rather than recomputing the entire analysis. In such settings, each language’s analysis must produce and refine function summaries that capture both local semantics and cross-language effects. Exchanging and reconciling these summaries typically requires multiple global iterations. A promising research direction is to employ differential transfer functions—recently explored by Kumar et al.~\cite{kumarIncrementalECOOP2025}—to compute fixed points incrementally, updating only the deltas in inter-language dataflow, thereby reducing recomputation cost.

\paragraph{Challenge: Generating the call-graph connecting three languages} Generating a precise inter-procedural call graph in three-language systems is not only a matter of adding the third language’s call edges. Computing the call graph also requires resolving cross-boundary dependencies and the processing of the APIs responsible for polyglot communication. One of them is \emph{eval(p)}, where $p$ is a function of guest language. This can be resolved by adding a caller and callee relation between the caller and callee method $p$. On the side of the other language, if the method $p$ calls a method on the shared variable $i$, say $i.m()$, then it also needs to add the caller and callee relation between the method $p$ and the method $m()$. The information about $i$ has to come from the host language, and then the analysis can proceed to compute the call graph. Extending to three languages will require a similar approach to compute the call graph.

Thus, analyzing three-language polyglot programs requires adapting existing static analysis techniques to the languages’ peculiarities. The analysis needs to run until it reaches a fixed point across all languages. It also needs to handle the APIs responsible for polyglot communication and compute the call graph. Moreover, it has to run incrementally to reduce the recomputation of the analysis.

\subsection{Summary based static analysis for multilingual communication through a common IR}

It is possible to leverage a common IR or any intermediate IR to perform the static analysis. For example, in the case of Java, Scala, and Kotlin analysis, the common IR can be a common intermediate language. After this, an intra-procedural analysis can be performed on the common IR by defining the transfer functions and computing those with respect to the program. Given a common intermediate representation (IR), the static analysis can be performed by defining the transfer functions for the analysis and computing the dataflow information. The dataflow information can be computed by propagating the dataflow information across the program. This is a standard approach for performing static analysis on a common IR.  An inter-procedural analysis is not as straightforward as an inter-procedural dataflow analysis. One of the problems is generating a call graph, a precursor to any inter-procedural dataflow analysis.

\paragraph{Challenge: Generating call-graphs}  Despite having a common IR, constructing the call graph requires a control flow analysis which is mutally recursive with data flow analysis. Even for the case of common IR, static analysis designers still need to be aware of the language-specific semantics to generate a precise call graph as the transfer fuctions change. Certain optimizations, such as those for typed programming languages, narrows down the search space by adding the relevant functions from the set of sub-types of a variable $i$ (class-hierarchy analysis). However, the presence of callbacks, which alters the control-flow, makes it non-trivial. The accuracy of the call-graph, in this case, depends on the accuracy of dataflow analysis.

In summary, generating the call graph for a polyglot system where multiple languages are compiled to a common IR is quite similar to generating the call graph for a program written in a single language. However, the static analysis designers need to be aware of the language-specific semantics to generate a precise call graph.  A summary-based dataflow analysis for these kinds of interlanguage communication is the same as the communication across the functions in the same language. The only difference is that the call graph is generated based on the semantics of each language. Once the calling targets are known for the function calls, then the dataflow is propagated from the callee to the caller and vice-versa, akin to a dataflow analysis in a single language. Thus, in this case, analyzing three-language polyglot programs requires adapting existing static analysis techniques to the peculiarities of the languages involved.

\section{Roadmap and Open Questions}
The next step toward scalable multilingual analysis is to understand how existing multilingual analyses behave once an additional language boundary is introduced. In three-language systems, constructing a precise inter-procedural call graph is not merely a matter of adding the third language’s edges. The analysis must process the APIs that enable polyglot communication and reason about cross-boundary dependencies: for example, call targets reachable in the entry language A can depend on dynamic choices made deeper in the chain (B→C), such as callbacks, factory patterns, or returned function objects. These interactions create multi-hop dependencies that traditional analyses cannot capture. Similar challenges arise in dataflow analysis, where intermediate languages may transform or aggregate information before propoagating it, leading to  imprecision and the potential for circular propagation of summaries across language boundaries.

We envision extending these concepts toward \emph{n-language analysis}, where interactions among multiple languages are modeled and analyzed within a single unified framework. Realizing this requires (1) sound modular summary representations that works across multiple language boundaries, (2) incremental fixed-point algorithms that converge efficiently as languages are added, and (3) intermediate representations that retain language provenance and semantics strongly enough to support shared reasoning. Ultimately, the goal is to move from customized tools made for two languages to \emph{language-agnostic} frameworks for reasoning over entire multilingual ecosystems.
Realizing these opens multiple research questions:
\begin{description}
	\item[\textbf{RQ1.}] How can modular summaries be propagated soundly across  language boundaries without complete re-analysis of each language? 
	\item[\textbf{RQ2.}] What incremental algorithms can ensure convergence when analyzing multi-language systems?
	\item[\textbf{RQ3.}] How does precision degrade across more languages, and how can this loss be measured or mitigated?
	\item[\textbf{RQ4.}] What abstractions or metadata must common intermediate representations include to preserve language provenance and semantics for n-language analysis?
\end{description}

\section{Discussion}

Our vision identifies opportunities and challenges in the analysis of multilingual programs. While we start with three-language systems, several limitations and open questions remain before this vision can be realized.

First, the paper provides conceptual models rather than implementations. The goal is to outline what makes three-language analysis different, not to propose a ready solution. Second, many of the challenges we present, such as summary propagation, also occur in two-language analyses. What changes in three-language scenarios is the interaction between boundaries: analyses must coordinate across multiple semantic interfaces, often forming dependency cycles that current frameworks cannot handle. Third, the field currently lacks formal models and benchmarks for studying n-language multilingual systems.

Finally, the long-term goal of n-language analysis raises scalability issues. As the number of languages increases, the complexity of analysis increases as well. Addressing these issues will require combining ideas from areas such as abstract interpretation, pointer analysis, and interprocedural analysis to make cross-language reasoning practical. Moreover, success in this area will depend on collaboration across research communities—bridging software engineering, programming languages, and AI-based analysis.

\section{Conclusion}
We outlined the landscape and challenges of three-language polyglot programs. Summary-based static analysis offers a scalable, precise approach for large programs. Extending this approach beyond two-language systems opens new opportunities for generalizing static analysis to multilingual ecosystems. Our vision aims to motivate systematic research toward sound, modular, and scalable analyses that can operate across multiple languages and runtimes.

\section*{Acknowledgements}
This work is %
funded by Digital Research Centre Denmark (DIREC), National Defence Technology Centre (NFC) and Danish Industry Foundation (project 2025-0766).

\begin{sloppypar}
\bibliographystyle{IEEEtran}
\bibliography{RelatedWork}
\end{sloppypar}
\end{document}